\RequirePackage{lineno}
\documentclass[aps,prl,twocolumn,superscriptaddress]{revtex4}

\usepackage{mathrsfs,amssymb,graphics,subfigure,threeparttable}
\usepackage{grap hicx}
\usepackage{amssymb}
\usepackage{enumerate}
\usepackage{amsmath}
\usepackage{fancyhdr}
\usepackage{bm}
\usepackage[squaren]{SIunits}
\usepackage{xcolor}
\usepackage{soul}

\begin{document}

%\preprint{}

\title{Generation of topologically complex three-dimensional electron beams in a plasma photocathode}


\author{Xinlu Xu}
\email[]{xuxinlu@slac.stanford.edu}
\affiliation{SLAC National Accelerator Laboratory, Menlo Park, CA 94025}
\author{Jorge Vieira}
\affiliation{GOLP/Instituto de Plasma e Fus\~ao Nuclear, Instituto Superior T\'ecnico, Universidade de Lisboa, 1049-001  Lisbon, Portugal}
\author{Mark J. Hogan}
\affiliation{SLAC National Accelerator Laboratory, Menlo Park, CA 94025}
\author{Chan Joshi}
\affiliation{Department of Electrical Engineering, University of California, Los Angeles, California 90095, USA}
\author{Warren B. Mori}
\affiliation{Department of Electrical Engineering, University of California, Los Angeles, California 90095, USA}
\affiliation{Department of Physics and Astronomy, University of California Los Angeles, Los Angeles, CA 90095, USA}

\date{\today}

\begin{abstract}
Laser-triggered ionization injection is a promising way of generating controllable high-quality electrons in plasma-based acceleration. We show that ionization injection of electrons into a fully nonlinear plasma wave wake using a laser pulse comprising of one or more Laguerre-Gaussian modes with combinations of spin and orbital angular momentum can generate exotic three-dimensional (3D) spatial distributions of high-quality relativistic electrons. The phase dependent residual momenta and initial positions of the ionized electrons are encoded into their final phase space distributions, leading to complex spatiotemporal structures. The structures are formed as a result of the transverse (betatron) and longitudinal (phase slippage and energy gain) dynamics of the electrons in the wake immediately after the electrons are injected. Theoretical analysis and 3D simulations verify this mapping process leads to the generation of these complex topological beams. These beams may trigger novel beam-plasma interactions as well as produce coherent radiation with orbital angular momentum when sent through a resonant undulator. 
\end{abstract}

%\keywords{}

\maketitle

%\section{\label{sec: Introduction}Introduction}

%\begin{figure}[bp]
%\includegraphics[width=0.5\textwidth]{fig1.pdf}
%\caption{\label{fig: }  }
%\end{figure}

%\linenumbers

Plasma-based acceleration (PBA) \cite{PhysRevLett.43.267, chen1985acceleration} is attractive because it can provide acceleration gradients in excess of $\giga\volt\per\centi\meter$. The last several decades have seen tremendous progress in PBA research \cite{joshi2020perspectives}, including the demonstration of high gradients \cite{blumenfeld2007energy, PhysRevLett.122.084801, adli2018acceleration} and the generation of electron beams suitable for applications, e.g., driving a compact free-electron-laser \cite{wang2021free} and advanced QED studies \cite{PhysRevX.8.011020, poder2018experimental}, producing bright and collimated X-rays \cite{RevModPhys.85.1} which have unlocked new research opportunities in high energy density science \cite{behm2020demonstration}, and imaging applications by providing high resolution three-dimensional images of biological samples \cite{cole2015laser}.

The production of high quality electron beams \cite{PhysRevLett.108.035001, PhysRevLett.111.015003, PhysRevLett.117.124801, PhysRevAccelBeams.20.111303, dalichaouch2020generating} has been instrumental for these advances. Various techniques were developed to controllably inject electrons into a relativistic plasma wake, such as downramp trapping \cite{PhysRevE.58.R5257, PhysRevLett.86.1011, PhysRevLett.100.215004, gonsalves2011tunable, PhysRevLett.110.185006, deng2019generation} and ionization injection \cite{chen2006electron, PhysRevLett.98.084801, PhysRevLett.104.025003, PhysRevLett.112.025001, deng2019generation}. Recent work on plasma cathodes has opened the possibility of generating femtosecond duration electron beams with MeV energy spreads, peak currents as high as hundreds of kA \cite{lundh2011few, emma2021terawatt} and normalized emittance $\epsilon_n$ as low as 10's of nm \cite{PhysRevLett.108.035001, PhysRevLett.111.015003, PhysRevLett.112.035003, PhysRevLett.112.125001, PhysRevSTAB.17.061301, PhysRevAccelBeams.20.111303, dalichaouch2020generating}. Besides having the potential to achieve unprecedented beam brightness, plasma cathodes can imprint multi-dimensional spatial structures onto the accelerated beams. For instance, several schemes purport to generate longitudinally bunched (1D) electrons \cite{lundh2013experimental, xu2016nanoscale, wenz2019dual, lumpkin2020coherent, xu2020generation} with potential to produce temporally coherent radiation. Electron rings observed in experiments from electron trapping within wake pockets created by sheath splitting \cite{pollock2015formation} and predicted to be generated in donut shaped wakefields driven by a relativistic high order Laguerre-Gaussian (LG) laser pulse ($a_L\equiv 8.6\times 10^{-10} \lambda[\micro\meter]I^{1/2} [\watt/\centi\meter^2] >1$) \cite{vieira2014nonlinear}, could be useful to clean the halo from heavy ion beams in conventional particle accelerators~\cite{PhysRevLett.107.084802}.

The topology of electron beams produced by self-injection in nonlinear plasma waves is both of fundamental \cite{mendoncca2009stimulated, vieira2016amplification, vieira2016high, vieira2018optical} and practical interest. Helically or sinusoidally modulated beams could be used to achieve superradiant emission \cite{RevModPhys.91.035003} of broadband X-rays in conventional and plasma-based light sources~\cite{vieira2021generalized}. Furthermore, relativistic beams with non-trivial topologies could potentially emit coherent radiation with orbital angular momentum (OAM) \cite{PhysRevA.45.8185, yao2011orbital}, beyond the visible spectrum~\cite{hemsing2013coherent}. Short-wavelength OAM (vortex) light is interesting because it can extend the OAM laser-matter interactions to the nanometer or even atomic scale thereby enable interesting applications in many fields \cite{van2007prediction, picon2010photoionization, rury2013examining, van2015interaction, jhajj2016spatiotemporal, hernandez2017generation}. However, current laser-plasma based schemes produce electron beams with spiral structures by transferring a large amount of angular momentum to them through the twisted wakefield driven by a relativistic light spring \cite{vieira2018optical} or the twisted electromagnetic fields of a super-intense laser ($a_L\sim 100$) with high-order LG mode \cite{liu2016generation,  ju2018manipulating, baumann2018electron}. These beams are characterized by large emittance and energy spread (almost continuous spectrum), and are therefore not suitable for producing coherent radiation.

In this Letter, we show through theory and supporting particle-in-cell (PIC) simulations that a plasma photocathode based on ionization injection into a nonlinear wakefield from non-relativistic lasers ($a_L\sim 0.1$) with combinations of spin and orbital angular momentum can generate high-quality electron beams with exotic 3D spiral spatial distributions. The intensity of the ionizing laser needs to be only marginally above ionization thresholds, $\sim 10^{17}~\mathrm{W/cm^2}$, thus such configuration is realizable with standard technology. It is now well appreciated that a circularly polarized (CP) laser carries what is referred to as spin angular momentum while a linearly polarized LG mode carries OAM \cite{PhysRevA.45.8185}. An OAM mode can  be described as $\vec{A}=- \mathrm{Re} (\sigma\vec{e}_x + i \hat{e}_y ) a_L c_p^{|l|} (r,\theta, z) \mathrm{e}^{i(l\theta - kz + \omega t)}$ where the angular momentum per photon is $l\hbar$ \cite{PhysRevA.45.8185}, $\omega$ and $k$ are the frequency and wavenumber of the laser, and $\sigma=1$ for right-handed CP and $-1$ for left-handed CP. The details for the complex functions $c_p^{|l|}$ are given in the supplement. 

\begin{figure}[bp]
\includegraphics[width=0.5\textwidth]{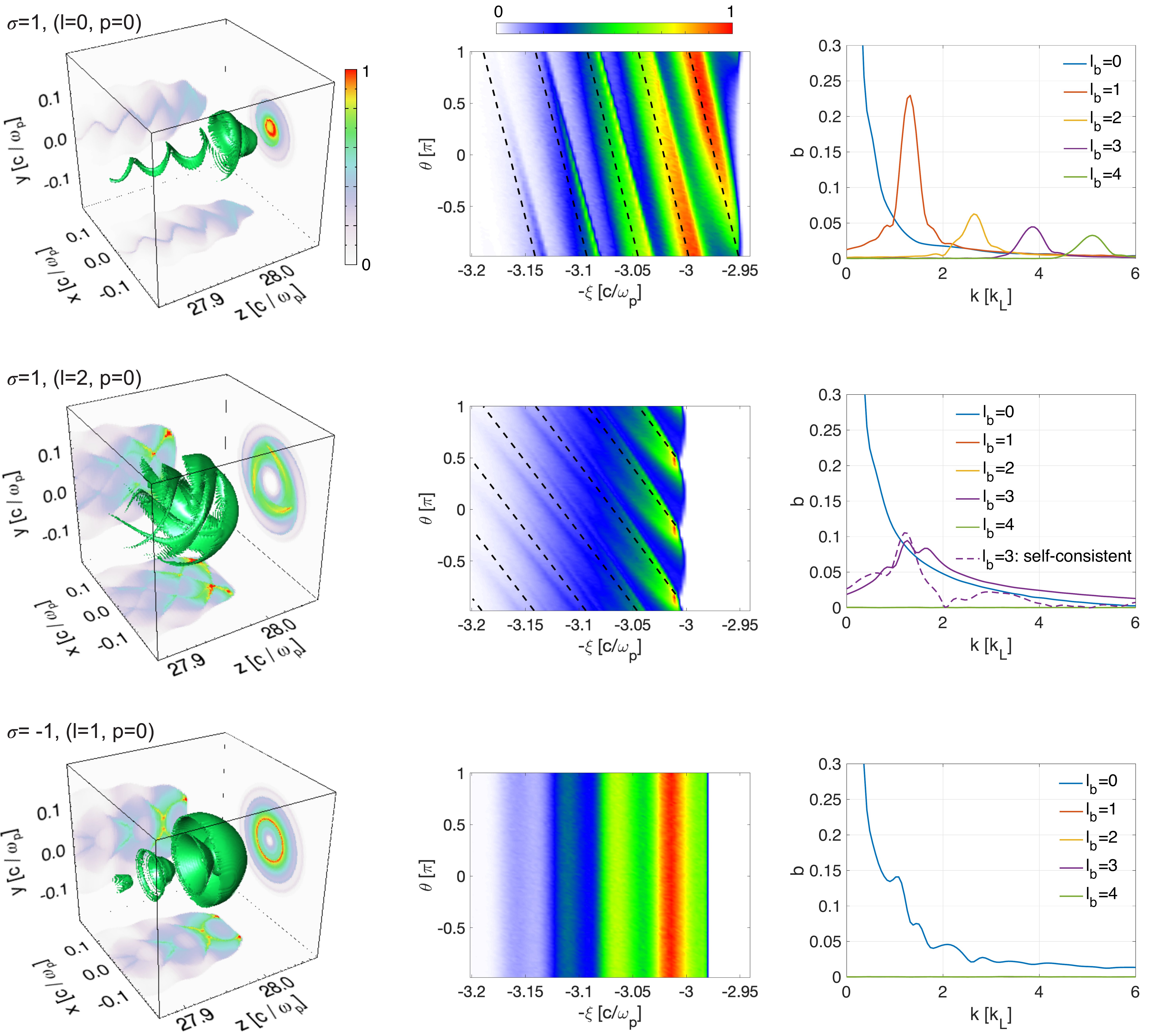}
\caption{\label{fig: results}  The structures of the injected electrons at $\omega_pt=31$. First column: Isosurface of the electron density and its projections on each plane. Second column: Normalized density distribution of the electrons in the $\theta-\xi$ plane. The black dashed lines represent the predications from Eq. (2). Third column: bunching factor. }
\end{figure}

In ionization injection electrons are born insider a fully blown out plasma wake through tunnel ionization of the electric field from one or more laser pulses. The 3D phase information of the laser(s) at the instant of ionization is imprinted onto the final electron distribution when the electrons become trapped after the laser pulse eventually overtakes them. The 3D spatial distributions evolve spatiotemporally within the wake as the electrons gain energy and phase slip longitudinally while executing betatron oscillations under the linear (transverse) focusing force of the ion column in the nonlinear wake. Although ionization injection has been extensively studied, there has been no investigation into how the spin and OAM of the laser is imprinted onto the angular momentum and distribution of the self-injected electrons. Here, we show that this concept permits designing beams with complex spatiotemporal distributions. The electrons generated in this new scheme can be characterized by small emittance ($\sim 100~\nano\meter$), small energy spread ($\sim 1.5~\mathrm{MeV}$), zero net angular momentum with a small spread ($\sim 100~\nano\meter\cdot mc $), and $\mathrm{kA}$ current that are suitable to produce high power short wavelength radiation with OAM, where $m$ is the electron mass and $c$ is the speed of light in vacuum. By using multiple laser pulses with different polarizations and LG modes, a beam with an axially varying spiral structure or multiple beams with different twisted structures can be produced.

To illustrate how this injection scheme can be used to generate spiraling and other complex 3D structured electron beams, we consider a fully blown out wake generated by an electron beam driver. We simulate the ionization injection by an appropriately delayed but co-moving ultrashort laser using non-evolving forces characteristic of nonlinear wakefields using the 3D PIC code OSIRIS \cite{fonseca2002high, xu2020numerical}. The forces for electrons with forward velocity $\beta_zc$ are $F_z=\frac{\xi}{2} m\omega_p^2, F_r= -[ \frac{r}{2}+(1-\beta_z)\frac{r}{2} ] m\omega_p^2$, where $\xi\equiv ct -z$ and  $r $ are the longitudinal and transverse coordinates and $\omega_p$ is the ambient plasma frequency. The newly ionized electrons are pushed in the prescribed wakefields, the laser field, and their own self-consistent fields. This significantly reduces the computational requirements as we only need to follow the injected beam particles. This approximation is well justified as the wake created by a highly relativistic electron beam driver evolves very slowly (hundreds of plasma periods for GeV-class beams) on the time scales of the injection process (several plasma periods). 

As shown in Fig. \ref{fig: results}, a CP 800 nm laser pulse with specified LG modes propagates through a mixture of majority hydrogen plasma with $n_p=1.74\times 10^{17}~\centi\meter^{-3}$ and minority He$^{1+}$ plasma with a density of $10^{-4}n_p$. The injected electrons are supplied via laser ionization of the He$^{1+}$ ions \cite{ADKionizationmodel1986}. The density of He$^{1+}$ is set to be low to minimize the space charge repulsion between the ionized electrons when they have low energies. The lasers are focused at $z=2\frac{c}{\omega_p}$ with a spot size $w_0=0.22\frac{c}{\omega_p}~(2.8~\micro\meter)$ and start at $z=-2\frac{c}{\omega_p}$ with a duration $\tau_{FWHM}=0.23\omega_p^{-1}~(9.8~\femto\second)$. The intensities of the pulses are adjusted to ensure similar injected charge in all cases: $a_L=0.085$ for $(l=0,p=0)$ while $a_L=0.14$ for other cases. The He$^{1+}$ plasma starts from $z=-2\frac{c}{\omega_p}$ to ensure that ionization within a Rayleigh length is included. 

Simulation results are presented in Fig. \ref{fig: results} (see supplemental material for details on the simulation parameters and more cases \cite{supplement}). We only consider $l\geq 0$ without loss of generality. Density isosurfaces of the trapped electrons are shown in the first column while their density distributions in the $(\theta,\xi)$ plane are shown in the second column where $\theta \equiv \mathrm{atan}2 (y, x)$ \cite{Atan2} is the angle in the transverse plane. As is clear, electrons with complex 3D structures are formed. When using a right-handed CP laser ($\sigma=1$) with a fundamental LG mode $(l=0,p=0)$, a single spiral beam (corkscrew) is produced; when $(l=2, p=0)$, three beamlets twist together to form a triple helical structure; when a left-handed CP laser ($\sigma=-1$) with $(l=1,p=0)$ is used, the spiral structure is absent altogether with the resulting beam forming a series of hollow shells. In the first two cases, the angle $\theta$ has an approximately linear dependence on the longitudinal position $\xi$ while this dependence is absent in the last case. The beams have $\epsilon_n \sim100~\nano\meter$ and $\sim 1.5$ MeV uncorrelated energy spread. Their energies at this time are $\sim20$ MeV and can be boosted to GeV-class in the following acceleration. 

We introduce a bunching factor $b(k,l_b)= \frac{1}{N} | \sum_{j=1}^N  \mathrm{exp}[i ( l_b\theta_j - k\xi_j) ]|$ to quantify the 3D structures, where $N$ is the number of the electrons. The results are shown in Fig. \ref{fig: results}, column 3. For the beam produced by the $l=0$ laser, the bunching factor is maximum at $(l_b=1, k\approx 1.3 k_L)$ and the beam is rich in spatial-harmonics, i.e., $(l_b=2, k\approx 2.6 k_L)$, $(l_b=3, k\approx 3.9 k_L)$ and $(l_b=4,  k\approx 5.2 k_L)$. When a laser with $l=2$ is used, the bunching factor is maximum at $(l_b=3, k\approx 1.5 k_L) $ while the harmonics are not distinctly present at this propagation time. A self-consistent simulation \cite{supplement} where an electron beam driver excites the wake and ionization of $n_{\mathrm{He}^{1+}}=0.05n_p$ provides the injected electrons is shown by the dashed line. The high current (2.5 kA) injected beam has a similar bunching factor for $l_b=3$ which validates the non-evolving force model and shows the structure is still formed even for kA currents. 

When the circular polarization direction of the $l=1$ laser is reversed from right-handed to left-handed, the bunching factor is zero for all $l_b\neq0$. Interestingly, the beams at this time are hollow which may be used to generate plasma wakes suitable for positron acceleration \cite{vieira2014nonlinear, PhysRevLett.115.195001}. 

To understand the simulation results, we propose a model for the dynamics of the ionized electrons. After being tunnel ionized, the electrons begin to move under the influence of the laser field and the plasma wakefield. When studying the longitudinal dynamics, the axial oscillations inside the laser pulse can be ignored since the energy gain from the low intensity and short laser is negligible compared to that due to the wake. The electrons are accelerated by the wake and then move nearly synchronously with the wake longitudinally. There is thus a longitudinal mapping between initial longitudinal position $\xi_i$ and the final nearly locked position $\xi$ as described in Ref. \cite{PhysRevLett.112.035003}, i.e., $k_p\xi \approx \sqrt{4 + (k_p\xi_i)^2 }$. 

The transverse motion of the injected electrons can be divided into two stages. In the first stage, after ionization, the electrons respond to the oscillating laser field. Since the field amplitude of the laser ($E_L \sim 10 \frac{mc\omega_p}{e}$) is typically much higher than the local value of the wakefield ($E_{wake, \perp}\sim 0.1\frac{mc\omega_p}{e}$), the electrons can be assumed to oscillate only under the influence of the laser field. The electrons are rapidly passed over by the ionizing laser since their longitudinal velocities are much less than $c$, and they then conduct betatron oscillations in the wakefield \cite{PhysRevLett.88.135004} . 

\begin{figure}[bp]
\includegraphics[width=0.5\textwidth]{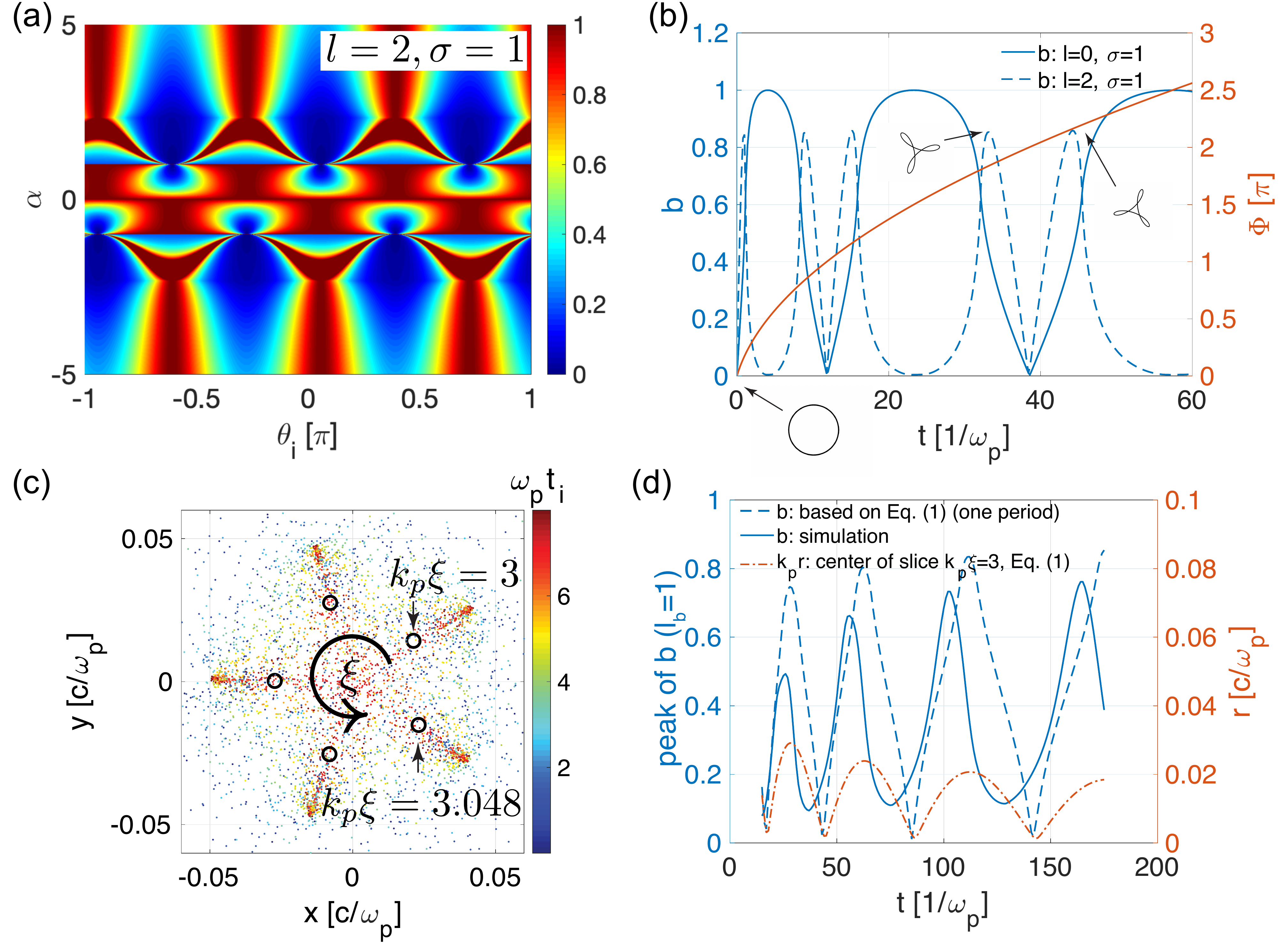}
\caption{\label{fig: theory} Theoretical analysis of the the twisted structure. (a) $g_{2,1}(\theta)$ with $k_p\xi_i=2.24$. The value at each $\alpha$ is normalized by 10 when $1\leq |\alpha| <2$ and its maximum for other $\alpha$. (b) The bunching factor ($k=0, l_b=l+\sigma$) and the betatron phase of the electrons ionized with $k_pr_i=0.1, a_i=0.1, k_p\xi_i=2.24$ and $t_i=0$. (c) $l=0, \sigma=1$: the distribution of the electrons and their centers (black circles) at $\omega_p t = 31$ in 5 different slices using Eq. (1). (d) The evolution of the peak of the bunching factor and the radial position of the center of the $k_p\xi=3$ slice. }
\end{figure}

The transverse canonical momentum $\vec{P}_\perp \equiv \vec{p}_\perp - e\vec{A}_\perp$ is conserved if a plane wave assumption is made. In the first stage, so long as $k_L w_0\gg1$ then canonical momentum remains approximately satisfied for LG modes. Thus, we assume the momenta of the electrons when they leave the laser pulse are equal to the vector potential at the instant of ionization. For a CP laser pulse of a single LG mode, the normalized vector potential in the two transverse directions are approximately $a_x \approx -\sigma a (r) \mathrm{cos} (l\theta - \hat{\xi}), a_y \approx a(r) \mathrm{sin} ( l \theta - \hat{\xi} )$, where $\hat{\xi} = k_L \xi$. We also assume that the transverse coordinates do not change during the transit time of the lasers \cite{xu2014low}. 

In the second stage, the electrons begin to respond to the wakefield. Assuming each electron experiences a constant acceleration gradient $E_z$ and its energy increases adiabatically, the asymptotic solution of the equation of motion is \cite{PhysRevLett.112.035003}
\begin{align}
&x \approx \left( \frac{1}{\gamma}\right)^{\frac{1}{4}} {x}_i \mathrm{cos}\Phi +   \left( \frac{4 }{\gamma}\right)^{\frac{1}{4}} \frac{ {p}_{xi} }{mck_p} \mathrm{sin} \Phi	\nonumber \\
&y \approx \left( \frac{1}{\gamma}\right)^{\frac{1}{4}} {y}_i \mathrm{cos}\Phi + \left( \frac{4 }{\gamma}\right)^{\frac{1}{4}} \frac{ {p}_{yi}}{mck_p}   \mathrm{sin} \Phi
\end{align}
where ${x}_i={r}_i \mathrm{cos}\theta_i, {y}_i = {r}_i \mathrm{sin}\theta_i, \frac{p_{xi}}{mc} \approx \sigma a_i \mathrm{cos}(l\theta_i - \hat{\xi}_i), \frac{p_{yi}}{mc} \approx -a_i \mathrm{sin}(l\theta_i - \hat{\xi}_i) $, and $a_i$ is the laser normalized vector potential when the electrons are ionized, $\Phi \approx \frac{\sqrt{2\gamma} - \sqrt{2}} {eE_z/(mc\omega_p)}$ is the betatron phase, $\gamma=1 + \frac{e E_z}{mc} (z-z_0)$ is the relativistic factor. Eq (1) can be used to determine $\theta=\mathrm{atan2}(y,x)$ \cite{supplement, Atan2}.

The laser pulse transfers part of its angular momentum (spin and orbital) to the ionized electrons as $L_z \equiv x_i p_{yi} - y_i p_{xi} \approx r_i a_i \mathrm{sin } [ (l + \sigma)\theta_i - \hat{ \xi}_i ] $ which is conserved when the electrons move inside an axisymmetric ion column. Since the electrons are ionized uniformly in $\theta_i$ by a CP laser, each slice possesses zero net angular momentum if $l+\sigma \neq 0$ with a small spread. On the other hand, when $l+\sigma=0$, each slice possesses a $\xi$-dependent angular momentum. The net transverse momentum of each slice is finite for $l=0$ and zero for $l\neq 0$. Thus, the center of each slice oscillates linearly for $l=0$ and stays at rest for other cases.  

To analyze the formation of these twisted structures we must map the initial spatial distribution of the ionized electrons to their present distribution. For electrons released at $r_i, \xi_i$ and $t_i$, the angular  distribution is $g_{l,\sigma}(\theta) = f(\theta_i)\left| \frac{\mathrm{d}\theta}{\mathrm{d}\theta_i}\right|^{-1} =  \left| \frac{ \alpha^2 +1  +  2\sigma \alpha \mathrm{cos} [(l+\sigma)\theta_i - \hat{\xi}_i] } { \alpha^2 - \sigma l  - \alpha  (l -\sigma )\mathrm{cos}  [(l+\sigma)\theta_i - \hat{\xi}_i]  } \right| $, where $f(\theta_i)=1, \alpha = \frac{k_p r_i}{\sqrt{2} a_i} \frac{1}{\tan \Phi}$ and $a_i$ is assumed to have a weak dependence on $r_i$. While $\theta(\theta_i)$ is known, there is no explicit expression for $\theta_i$ as a function of $\theta$. However, we can still make some useful observations. Clearly if $l\geq 1$, then if $\alpha = 0$ or $\infty$, $g_{l,\sigma} (\theta)$ is a constant. The variable $\alpha$ evolves as the particles are accelerated and $\alpha=0~(\infty)$ when $\Phi=n\pi+\frac{\pi}{2}~(n\pi)$ where $n$ is an integer. We next discuss some behaviors for a right-handed CP laser: $g_{l\geq 1, 1}(\theta)$ achieves its maxima at $\theta_i = \frac{2n\pi + \hat{\xi}_i}{l+1} $ for $\alpha\geq l$ or $0\leq \alpha<1$, at $\theta_i = \frac{(2n+1)\pi + \hat{\xi}_i}{l+1} $ for $\alpha< -l$ or $-1 \leq \alpha < 0$, and at a $\theta_i$ when the denominator vanishes for $1\leq |\alpha| < l$. Fig. \ref{fig: theory} (a) shows the dependence of $g_{2,1}$ on $\alpha$ and $\theta_i$. The angles $\theta_i = \frac{2n\pi + \hat{\xi}_i}{l+1}$ or $\frac{(2n+1)\pi + \hat{\xi}_i}{l+1}$ are mapped to $\theta=\theta_i$ when $(\alpha + 1)\sin \Phi >0$ and $\theta=\theta_i + \pi$ when $(\alpha+1)\sin\Phi<0$. Thus at some betatron phases, these electrons are concentrated at $l+\sigma$ equally spaced angles which depend linearly on $\xi_i$. This concentration has a quasi-period of the betatron phase, $\frac{\pi}{2}$. More details can be found in the supplemental material. For these special angles, $g_{l,\sigma}(\theta)$ is then known. 

For $l=0$, the dynamics is different, i.e., when $\alpha=0~(\Phi=n\pi+\frac{\pi}{2})$, then from Eq. 1 it can be seen that $r=\gamma^{1/4}\frac{\sqrt{2}a_i}{k_p}$ and $\theta=\hat{\xi}_i ~(\sin \Phi>0)$ or $\hat{\xi}_i+\pi~(\sin\Phi<0)$. This indicates a single spiral beam is formed. Thus the angular distribution has a quasi-period of the betatron phase, $\pi$. The bunching factor of the electrons for $l=0$ and $l=2$ in Fig. \ref{fig: theory}(b) confirms the quasi-periodic behavior of the angle distribution. The insets show the distribution at $\omega_pt=0, 33$ and $44$ for $l=2$.

Physically the electrons in each slice form $ l + \sigma$ beamlets and the center of each one conducts linearly oscillations with a $\xi_i$-dependent angle. The twisted structures do not rotate, they only flip when the centers cross the origin.

Electrons are ionized at different $r_i$, which complicates how electrons are distributed in $\theta$. However, there are always betatron phases $\Phi$ where electrons are concentrated at $l+\sigma$ angles due to evolution of $\alpha\propto \frac{1}{\tan \Phi}$. Furthermore, due to longitudinal mixing \cite{PhysRevLett.112.035003}, i.e., one slice contains electrons ionized at different times $t_i$), which leads to a spread of the phase, which blurs the twisted structure of the beam. The betatron phase grows slower as the electrons gain energy, thus the rms spread of the phase decreases as $\sigma_\Phi \approx \frac{k_pL_{inj}}{\sqrt{12}}\frac{1}{\sqrt{2\gamma}}$ \cite{PhysRevLett.112.035003}, where $L_{inj}$ is the distance over which ionization occurs. In the simulations presented here  $k_pL_{inj}\sim 8$. As a result, the amplitude of the oscillations of $b$ increases monotonically during the acceleration as shown in Fig. \ref{fig: theory}(d). Fig. \ref{fig: theory}(c) shows the distribution of the electrons with different $r_i$ and $t_i$ at 5 slices ($k_p\xi_i=2.236, 2.249, 2,262, 2.287, 2.3$ which correspond to $k_p\xi=3, 3.01, 3.019, 3.029, 3.038, 3.048$ based on the longitudinal mapping). For each $k_p\xi_i$, the values of $r_i$ and $t_i$ are consistent with what is seen in simulations. The concentration at certain angles is clearly seen. In Fig. \ref{fig: theory}(d), the long-term behavior of both $b$ and the $r$ of the center shows the oscillations and the increase of $b$ with time is seen.

When the laser polarization is left-handed CP, similar conclusions can be obtained when $l\neq 1$. However, in this case when $l=1$, $g_{1,-1}(\theta) $ does not depend on $\theta$, which indicates that the electrons are distributed uniformly in $\theta$, which is consistent with the third row of Fig. \ref{fig: results}. When a linearly polarized LG laser mode is used, the intrinsic discretization when ionizing the electrons can produced a longitudinally bunched structure \cite{xu2016nanoscale, PhysRevLett.125.014801}. 

\begin{figure}[bp]
\includegraphics[width=0.5\textwidth]{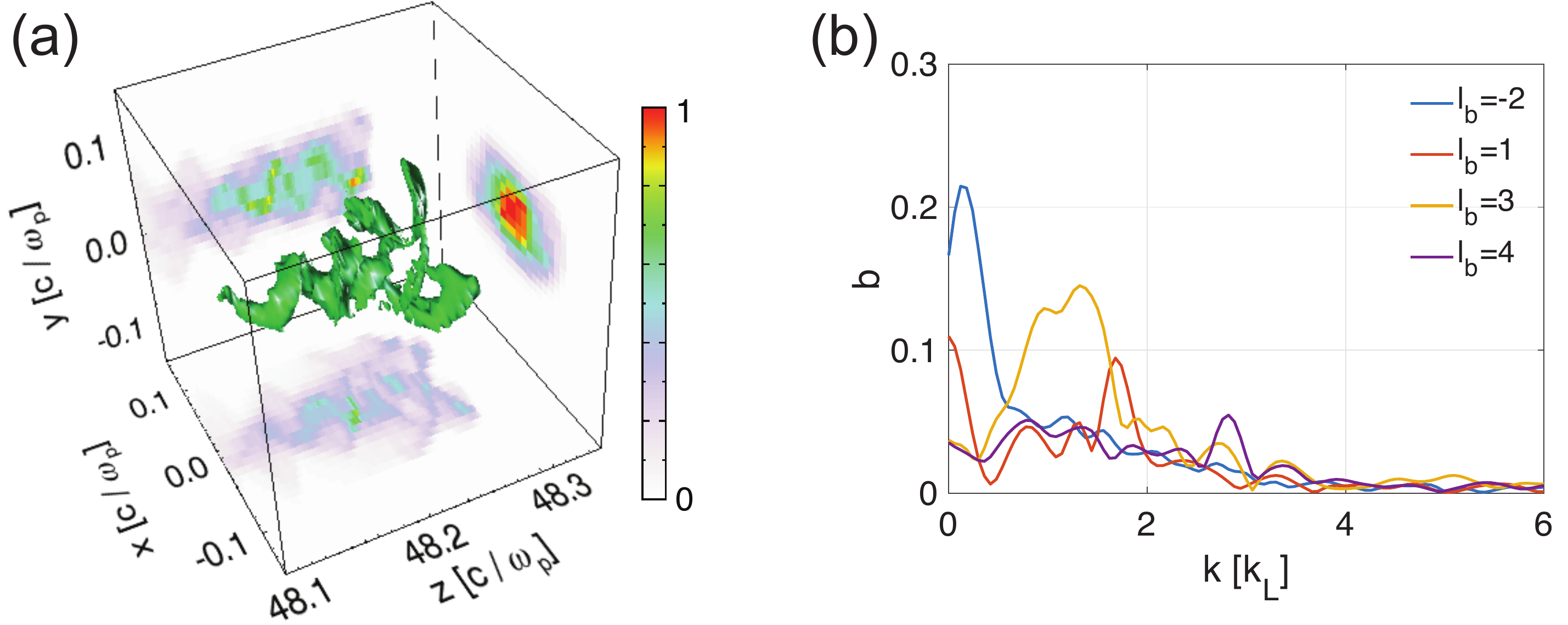}
\caption{\label{fig: two pulses}   The structure of the injected electrons when two right-handed CP laser pulses with $(l=0,p=0)$ and $(l=2,p=0)$ are used. (a) Density isosurface and its projections to each plane; (b) the bunching factor. The profiles of the pulses are as same as in Fig. (1).}
\end{figure}

Based on the aforementioned longitudinal mapping \cite{PhysRevLett.112.035003}, the electrons are concentrated at angles that can be written as a function of their positions $\xi$ after injection. For example, when $\alpha < - l$ and $(\alpha +1 )\mathrm{sin} \Phi >0$ ($\theta_i$ where $g_{l,\sigma}$ is maximum is mapped to $\theta=\theta_i$), 
\begin{align}
\theta \approx \frac{2 n + 1}{| l + \sigma|} \pi + \frac{ (k_L/k_p) \sqrt{ (k_p\xi)^2-4 }}{|l + \sigma|} 
\end{align}
Eq. (2) indicates there are $| l + \sigma|$ beamlets spiraling around each other. The comparison between the angles at which electrons are concentrated in the $\theta-\xi$ plane and those predicted by Eq. (2) (dashed lines) are shown in the second column of Fig. \ref{fig: results}. Good agreement is obtained.

The mapping from the phase distribution of LG-CP lasers to the 3D structure of the injected electrons was discussed above. This mapping can be extended to lasers with arbitrary phase distribution and electrons with more complicated structures being produced. Here we show an example: two right-handed CP laser pulses, one with $(l=0, p=0), a_L=0.057$ and the other with $(l=2,p=0), a_L=0.106$, co-propagate into the nonlinear wake driven by an 1 GeV electron beam with 19kA peak current and the centers of these two lasers are $k_p\xi = 2.5$ and $k_p \xi = 2.3$ respectively. The structure of the injected electrons gradually evolves from 3 beamlets at the head of the beam to 1 beamlet at the tail of the beam [Fig. \ref{fig: two pulses}(a)]. The bunching factor achieves the maximum at $(l_b=1, k\sim 1.7k_L)$ and their harmonics from the $l=0$ laser, and at $(l_b=3, k\sim 1.3k_L)$ from the $l=2$ laser. Additionally the bunching factor achieves the maximum at $(l_b=-2, k\approx 0.12 k_L)$ and $(l_b=4, k\approx 2.8 k_L)$ which is due to the mutual interactions (beating) between these two laser pulses. The electron beam has a 1.2 kA peak current, a 105 (86) nm emittance and a 1.3 MeV uncorrelated energy spread. By using laser pulses that have different wavelengths, modes, angles, and delays one can produce exotic 3D structured electron beams.

We point out that in contrast to previous work \cite{liu2016generation, ju2018manipulating, baumann2018electron, vieira2018optical} our work proposes a new scheme where the electrons form twisted structures while conducting the betatron motion in the linear fields of an ion column. These beams acquires zero net angular momentum with finite spread from the laser pulse. This is different from other work where the beam possesses a significant amount of angular momentum \cite{liu2016generation, vieira2018optical, ju2018manipulating, baumann2018electron} and the spiral motion of ions is needed to conserve the angular momentum. The beams produced in this work are characterized by small emittance ($\sim 100~\nano\meter$), small energy spread ($\leq 1.5~\mathrm{MeV}$) and several $\mathrm{kA}$ current. These beams are suitable to produce high power coherent radiation with orbital angular momentum (OAM) from ultraviolet to X-ray if they are boosted to high energy and propagate through a magnetic undulator \cite{hemsing2013coherent}.

\begin{acknowledgments}
This work was supported by the U.S. Department of Energy under contract number DE-AC02-76SF00515, US National Science Foundation grant number 1806046 and the US Department of Energy grant number DE-SC0010064 and a SciDAC FNAL subcontract 644405. The simulations were performed on the resources of the National Energy Research Scientific Computing Center (NERSC), a U.S. Department of Energy Office of Science User Facility located at Lawrence Berkeley National Laboratory, through an ALCC grant. J.V. acknowledges the support of FCT (Portugal) grant no. SFRH/IF/01635/2015.
\end{acknowledgments}

\bibliography{refs_xinlu}

\end{document}